\documentclass{article}

\usepackage[preprint]{neurips_2020}

\usepackage[utf8]{inputenc} 
\usepackage[T1]{fontenc}    
\usepackage{url}            
\usepackage{booktabs}       
\usepackage{amsfonts}       
\usepackage{nicefrac}       
\usepackage{microtype}      

\usepackage{tikz}
\usepackage{comment}
\usepackage{amsmath,amssymb} 
\usepackage{color}

\usepackage{times}

\usepackage{soul}
\usepackage{url}
\usepackage[utf8]{inputenc}
\usepackage[small]{caption}
\usepackage{graphicx}
\usepackage{booktabs}
\usepackage{amssymb}
\usepackage{comment}
\usepackage{amsthm}

\usepackage{bm}
\usepackage{float}
\usepackage{graphicx}
\usepackage{subfigure}
\usepackage{xcolor}
\usepackage{multirow}
\usepackage{array}
\newcolumntype{C}[1]{>{\centering\let\newline\\\arraybackslash\hspace{0pt}}m{#1}}
\title{DURRNet: Deep Unfolded Single Image Reflection Removal Network}

\author{%
    Jun-Jie Huang, Tianrui Liu, Zhixiong Yang, Shaojing Fu, Wentao Zhao, and Pier Luigi Dragotti
}

\begin{document}

\maketitle

\begin{abstract}
Single image reflection removal problem aims to divide a reflection-contaminated image into a transmission image and a reflection image. It is a canonical blind source separation problem and is highly ill-posed.
In this paper, we present a novel deep architecture called deep unfolded single image reflection removal network (DURRNet) which makes an attempt to combine the best features from model-based and learning-based paradigms and therefore leads to a more interpretable deep architecture.
Specifically, we first propose a model-based optimization with transform-based exclusion prior and then design an iterative algorithm with simple closed-form solutions for solving each sub-problems. With the deep unrolling technique, we build the DURRNet with ProxNets to model natural image priors and ProxInvNets which are constructed with invertible networks to impose the exclusion prior. Comprehensive experimental results on commonly used datasets demonstrate that the proposed DURRNet achieves state-of-the-art results both visually and quantitatively.

\end{abstract}

\section{Introduction}
\label{sec:intro}
Single image reflection removal (SIRR) is a typical blind image separation problem. It aims to decompose an image, which is captured through a glass and is associated with reflections, into a transmission image and a reflection image. The transmission image refers to the image content of the target scene on the other side of the glass, and the reflection image refers to the image content from another scene reflected by the glass. 
This is a highly ill-posed problem and requires high-level understanding of the scene. 

A reflection-contaminated color image $\mathbf{I} \in \mathbb{R}_{+}^{W \times H \times 3}$ is usually assumed to be a linear combination of a transmission image $\mathbf{T} \in \mathbb{R}_{+}^{W \times H \times 3}$ and a reflection image $\mathbf{R} \in \mathbb{R}_{+}^{W \times H \times 3}$, \textit{i.e.,} $\mathbf{I} = \mathbf{T} + \mathbf{R}$, where $W$ and $H$ are the width and height of the image, respectively. Decomposing $\mathbf{I}$ into $\mathbf{T}$ and $\mathbf{R}$ is a highly ill-posed problem since there are infinite number of feasible decomposition in the form of $\mathbf{I} = \left(\mathbf{T} + \mathbf{Q} \right) + \left( \mathbf{R} -\mathbf{Q} \right)$, where $\mathbf{Q}$ is 
the shared image content between $\mathbf{T}$ and $\mathbf{R}$. The purpose of image reflection removal is therefore to minimize the shared image contents on the decomposed images, and at the same time maintain the natural aspect of the estimated images.

In order to perform effective reflection removal, suitable priors should be exploited to constrain the problem effectively. Model-based methods~\cite{levin2004separating,levin2007user,li2014single,ghost_cues_2015,reflect_suppression2017,fast_convex_2019} formulate the image reflection removal problem as an optimization problem with explicitly defined image priors, for example, the gradient sparsity prior. Model-based methods lead to highly interpretable mathematical formulations and optimization algorithms though the end result may not be satisfactory when strong and complex reflections are present. 
On the other hand, methods based on deep learning~\cite{generic_smooth_2017,wan2017benchmarking,perceptual_loss_2018,yang2018seeing,beyond_linear_2019,wei2019single,cascaded_refine_2020} design task specific deep network structures and loss functions to exploit data-driven priors. These priors can be learned from large-scale real training data or from the generation of faithful synthetic training data.
However, the deep-learning based methods are difficult to interpret and a more principled approach to design the network structures is needed.

\begin{figure}[t]
\begin{centering}
\center
\includegraphics[width=0.85\columnwidth]{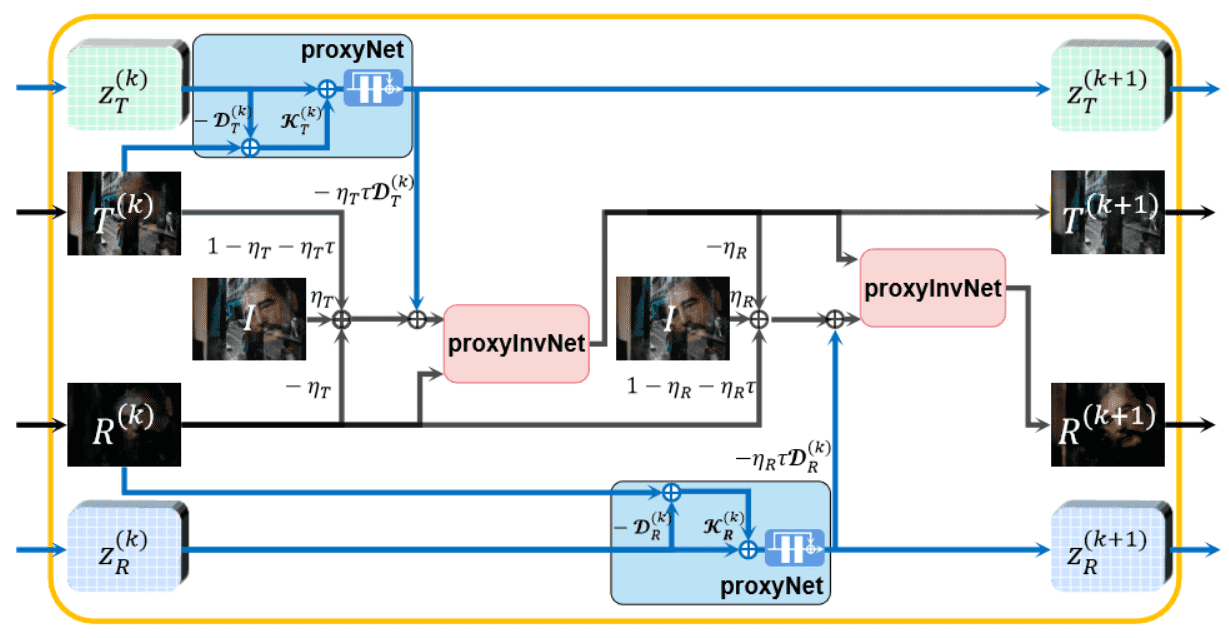}
\par\end{centering}
\caption{The proposed Deep Unfolded Reflection Removal Layer (DURRLayer) based on deep unfolding. It consists of a transmission estimation network and a reflection estimation network. For each estimation network, a proxyNet updates the features and the proxInvNet imposes exclusion condition on two estimated images.
}
\label{fig:DUSepLayer}
\end{figure}

In this paper, we propose a model-inspired deep network architecture for the image separation task using deep unrolling technique.
We first formulate the single image reflection removal problem as a convolutional sparse coding problem with sparsity priors and an exclusion prior, then we propose an iterative algorithm based on proximal gradient descent to solve the problem. By using the unfolding technique, we unroll an iteration of the proposed iterative algorithm into a Deep Unfolded Reflection Removal Layer (DURRLayer) as shown in Fig. \ref{fig:DUSepLayer}. A model-driven multi-scale Deep Unfolded Reflection Removal Network (DURRNet) is then constructed with DURRLayers in a multi-resolution fashion. Facilitated by the model-driven deep network structure, the proposed DURRNet is not only more interpretable, but also achieves high quality reflection removal results. 

The contribution of this paper is three-fold:
\begin{itemize}
    \item We propose a single image reflection removal convolutional sparse coding model by exploiting the formation model of reflection-contaminated image and a transform-based exclusion loss. Based on proximal gradient descent, we propose an iterative algorithm with simple computations.
    
    \item Based on the proposed iterative algorithm, we design a new deep network architecture for single image reflection removal by unrolling the algorithm into a deep network with learnable parameters. The proposed DURRNet consists of multiple scales of DURRLayers which has an exact step-by-step relationship with the corresponding optimization algorithm, therefore, is of high interpretability.
    
    \item Through extensive experiments, we demonstrate that the proposed DURRNet is able to achieve effective single image reflection removal and obtains highly competitive results compared to both the model-based and deep-learning based single image removal methods. 
    
\end{itemize}

The rest of the paper is organized as follows: Section \ref{sec:background} reviews the related single image removal methods and algorithm unfolding. Section~\ref{sec:method} presents the model formulation, optimization algorithm design and the deep network architecture of the proposed DURRNet. Section~\ref{sec:results} demonstrates the experimental results and comparisons. Section~\ref{sec:conclusion} concludes the paper.

\section{Related Works}
\label{sec:background}



\noindent \textbf{Model-based SIRR Methods} \cite{levin2004separating,levin2007user,li2014single,ghost_cues_2015,reflect_suppression2017,fast_convex_2019} formulate the image reflection removal problem as an optimization problem and solve it with optimization tools.
The gradient sparsity prior of natural images has been exploited in~\cite{levin2004separating,levin2007user} to obtain decomposition with minimal edges and local features. The relative smoothness prior has been proposed in~\cite{li2014single} since the reflected image is usually more blurred.
In~\cite{fast_convex_2019}, a convex model which implies a partial differential equation with gradient thresholding is used to suppress the reflection from a single input image. The Laplacian fidelity prior and the $l_0$ gradient sparsity prior have been used in~\cite{reflect_suppression2017} to formulate the optimization problem for reflection suppression. In~\cite{ghost_cues_2015}, Gaussian Mixture Model (GMM) has been applied for modelling patch prior to exploit the ghosting effects on reflection. 

 
\noindent \textbf{Deep-Learning-based SIRR Methods} ~\cite{generic_smooth_2017,wan2017benchmarking,perceptual_loss_2018,yang2018seeing,beyond_linear_2019,wei2019single,cascaded_refine_2020,hu2021trash} solve the reflection removal problem by designing proper deep network architectures, loss functions and exploiting external real or synthetically generated training datasets. 
The Cascaded Edge and Image Learning Network (CEILNet)~\cite{generic_smooth_2017} consists of two cascaded CNN networks, \textit{i.e.,} E-CNN and I-CNN for edge prediction and image reconstruction, respectively. In~\cite{perceptual_loss_2018}, exclusion loss, perceptual loss and adversarial loss are proposed to regularize the learning of the reflection separation network. In~\cite{yang2018seeing}, a bidirectional network (BDN) which consists of a cascaded deep network has been proposed to estimate the reflection image and use it to improve the estimation of the transmission image.
ERRNet~\cite{wei2019single} proposes to utilize misaligned training data with an alignment-invariant loss. In~\cite{cascaded_refine_2020}, an Iterative Boost Convolutional LSTM Network (IBCLN) has been proposed to progressively separate the reflection-contaminated image into two image layers. In~\cite{hu2021trash}, a dual-stream decomposition network has been proposed to enable information exchange at different branches and achieved state-of-the-art single image reflection removal performance.

\noindent \textbf{Deep Unfolding}~\cite{monga2021algorithm} aims to merge model-based and deep-learning based approaches for solving inverse problems (\textit{e.g.,} image restoration problems). The general idea is to design an iterative algorithm for the problem at hand and then convert certain steps of the iterative algorithm into learnable parameters. 
In the seminal work \cite{gregor2010learning}, Gregor and LeCun proposed to convert the iterative shrinkage-thresholding algorithm (ISTA) into a deep network by setting the dictionaries in ISTA as learnable parameters. In \cite{yang2016deep}, ADMM-Net has been proposed to unfold the Alternating Direction Method of Multipliers (ADMM) algorithm for compressive sensing Magnetic Resonance Imaging (MRI) reconstruction.
In~\cite{zhang2020deep}, a deep unfolding network for single image super-resolution has been proposed by unfolding Maximum-a-Posteriori (MAP) formulation via a half-quadratic splitting algorithm and interpreting the prior term as a denoiser. A Deep Unrolling for Blind Deblurring (DUBLID) network~\cite{li2020efficient} unfolds a total variation based blind deconvolution algorithm and contains a very small number of learnable parameters.
In~\cite{model_driven_rain_2020}, Wang \textit{et al.} proposed a model-inspired rain removal deep unfolding network based on proximal gradient descent to simplify computations.  Recently, Pu \textit{et al.}~\cite{pu2022mixed} proposed a self-supervised deep unfolding network for separating X-Ray images of Artworks.

\section{Proposed Method}
\label{sec:method}

In this section, we will first introduce the proposed model-based optimization formulation for single image reflection removal and then we solve the optimization using an iterative algorithm based on proximal gradient descent. Finally we present the proposed Deep Unfolded Reflection Removal Network (DURRNet) architecture based on the designed iterative algorithm and detail the training strategy.

\subsection{Model Formulation}
\label{sec:formulation}

A reflection-contaminated color image $\mathbf{I} \in \mathbb{R}_{+}^{W \times H \times 3}$ can be expressed as a linear combination of a transmission image and a reflection image~\cite{reflect_suppression2017}. 
Therefore, we can represent the observed reflection-contaminated color image as $\mathbf{I} = \mathbf{T} + \mathbf{R}$,
where $\mathbf{T} \in \mathbb{R}_{+}^{W \times H \times 3}$ and $\mathbf{R} \in \mathbb{R}_{+}^{W \times H \times 3}$ are the transmission image and the reflection image, respectively. $W$ and $H$ are the width and height of the image.

The reflection image is usually considered as a blurred version of the reflected scene due to the effect of the glass.
With different characteristics, $\mathbf{T}$ and $\mathbf{R}$ are assumed to have two different representations over a transmission dictionary $\mathbf{D}_T$ and a reflection dictionary $\mathbf{D}_R$, respectively. 
Based on the Convolutional Sparse Coding (CSC) model~\cite{papyan2017convolutional,bristow2013fast}, we propose to formulate the reflection removal problem as:
\begin{equation}
    \begin{aligned}
        \underset{\mathbf{z}_T, \mathbf{z}_R}{\min} & \frac{1}{2} \Vert \mathbf{I} - \sum_{i=1}^N \mathbf{D}_T^i \otimes \mathbf{z}_T^i - \sum_{j=1}^N \mathbf{D}_R^j \otimes \mathbf{z}_R^j \Vert_F^2 + \lambda_T p_T(\mathbf{z}_T) + \lambda_R p_R(\mathbf{z}_R),
    \end{aligned}
    \label{eq:model-csc}
\end{equation}
where $\mathbf{D}_T = [\mathbf{D}_T^1, \cdots, \mathbf{D}_T^N]$ and $\mathbf{D}_R = [\mathbf{D}_R^1, \cdots, \mathbf{D}_R^N]$ are the transmission convolutional dictionary and the reflection dictionary and $\mathbf{z}_T = [\mathbf{z}_T^1, \cdots, \mathbf{z}_T^N]$ and $\mathbf{z}_R = [\mathbf{z}_R^1, \cdots, \mathbf{z}_R^N]$ are the features corresponding to $\mathbf{T}$ and $\mathbf{R}$, respectively. Here $\otimes$ denotes the convolution operator and $N$ is the number of filters. Moreover, $\lambda_T$, $\lambda_R$ are regularization parameters, and $p_T(\cdot)$ and $p_R(\cdot)$ represents the prior term for the feature of $\mathbf{T}$ and $\mathbf{R}$, respectively.

The exclusion loss~\cite{perceptual_loss_2018} is based on the idea that if two images do not contain shared contents, then their edges and their contours will only overlap in a small region.
In~\cite{perceptual_loss_2018}, the exclusion loss is applied as a training loss function to facilitate the training of the image reflection network. It measures the degree of edge overlapping of two images in a multi-scale manner and can be expressed as:
\begin{align}
    \label{eq:excl1} \mathcal{L}_{\text{e}} = \sum_{j=1}^J || \Psi ( f^{\downarrow j}(\mathbf{T}), f^{\downarrow j}(\mathbf{R}) ) ||_F, 
\end{align}
where $\Psi (\mathbf{T}, \mathbf{R}) =\tanh(\beta_T |\nabla \mathbf{T}|) \odot \tanh(\beta_R |\nabla  \mathbf{R}|)$, $\beta_T$ and $\beta_R$ are normalization factors, moreover, $\odot$ denotes element-wise multiplication, $\nabla \mathbf{T}$ and $\nabla \mathbf{R}$ denote the gradients of $\mathbf{T}$ and $\mathbf{R}$, respectively. Finally, $f^{\downarrow j} (\cdot)$ denotes the downsampling operation by a factor $2^{j-1}$ with bilinear interpolation.


In our model, we aim to explicitly include the exclusion constraint into the optimization formulation for reflection removal, however, Eq. (\ref{eq:excl1}) does not lead to easy to compute solutions. 
Inspired by~\cite{kamilov2016parallel} which proposed a proximal-gradient algorithm for minimizing Total Variation regularized least-squares cost functional, a transform-based exclusion loss has been proposed in~\cite{pu2022mixed}:
\begin{equation}
    \mathcal{L}_{\text{te}}( \mathbf{T},  \mathbf{R}) = \sum_{m=1}^{M} \Vert \left( \mathbf{W}_m \otimes \mathbf{T} \right) \odot \left( \mathbf{W}_m \otimes \mathbf{R} \right) \Vert_1,
    \label{eq:excl_new}
\end{equation}
where $\mathbf{W} = [\mathbf{W}_1, \cdots, \mathbf{W}_M]$ denotes the high-pass filters of a transform with $\mathbf{W}_m$ being the $m$-th filter. 
This new formulation uses high-pass filters of a transform to extract high-frequency information from the image and measures the element-wise correlation between each pair of “edge” images in $l_1$ norm. 
This enables simple closed-form solution for the optimization problem. 

Based on Eq. (\ref{eq:model-csc}) and Eq. (\ref{eq:excl_new}), we propose to formulate the reflection removal problem as a convolutional sparse coding problem:
\begin{equation}
    \begin{aligned}
        \underset{\mathbf{z}_T, \mathbf{z}_R}{\min} & \frac{1}{2} \Vert \mathbf{I} - \mathbf{D}_T \otimes \mathbf{z}_T - \mathbf{D}_R \otimes \mathbf{z}_R \Vert_F^2 + \lambda_T p_T(\mathbf{z}_T) + \lambda_R p_R(\mathbf{z}_R) \\
        &+ \kappa \mathcal{L}_{\text{te}} (\mathbf{D}_T \otimes \mathbf{z}_T, \mathbf{D}_R \otimes \mathbf{z}_R),
    \end{aligned}
    \label{eq:model}
\end{equation}
where with a slight abuse of notation, we denote $\mathbf{D}_T \otimes \mathbf{z}_T = \sum_{i=1}^N \mathbf{D}_T^i \otimes \mathbf{z}_T^i$ and $\mathbf{D}_R \otimes \mathbf{z}_R = \sum_{i=1}^N \mathbf{D}_R^i \otimes \mathbf{z}_R^i$, and $\kappa$ is the regularization parameter for the exclusion term.


In Eq. (\ref{eq:model}), the transmission image and the reflection image are modelled as a linear combination of atoms from the transmission dictionary and the reflection dictionary; the data fidelity term ensures the estimated transmission image and the reflection image contain sufficient information of the observed image; the two prior terms, $p_T(\mathbf{z}_T)$ and $p_R(\mathbf{z}_R)$ regularize the features for the transmission and the reflection image, and the transform-based exclusion term $\mathcal{L}_{\text{te}}$ is used to further facilitate the separation of image contents on the two images.


\subsection{Optimization Algorithm}
\label{sec:algorithm}

Based on the model formulation defined in Eq. (\ref{eq:model}), in this section, we design an algorithm which solves iteratively simpler sub-problems for which we can provide close-form solutions. 
Since the features $\bm{z}_T$ and $\bm{z}_R$ appear in the data fidelity term, the prior terms and the exclusion terms, it is difficult to optimize all these terms jointly. Therefore, we introduce two auxiliary parameters $\hat{\mathbf{T}} = \mathbf{D}_T \otimes \mathbf{z}_T$ and $\hat{\mathbf{R}} = \mathbf{D}_R \otimes \mathbf{z}_R$. With Half-Quadratic Splitting (HQS) algorithm, Eq. (\ref{eq:model}) can then be reformulated as:
\begin{equation}
    \begin{aligned}
        \underset{\mathbf{z}_T, \mathbf{z}_R, \hat{\mathbf{T}}, \hat{\mathbf{R}}}{\min} & \frac{1}{2} \Vert \mathbf{I} - \hat{\mathbf{T}}- \hat{\mathbf{R}} \Vert_F^2  +\frac{\tau}{2} \Vert \hat{\mathbf{T}} - \mathbf{D}_T \otimes \mathbf{z}_T \Vert_F^2 +\frac{\tau}{2} \Vert \hat{\mathbf{R}} - \mathbf{D}_R \otimes \mathbf{z}_R \Vert_F^2 \\
         &+ \lambda_T p_T(\mathbf{z}_T)+ \lambda_R p_R(\mathbf{z}_R) + \kappa \sum_{m=1}^{M} \Vert ( \mathbf{W}_m \otimes \hat{\mathbf{T}} ) \odot ( \mathbf{W}_m \otimes \hat{\mathbf{R}}) \Vert_1,
    \end{aligned}
    \label{eq:model2}
\end{equation}
where $\tau$ is a regularization parameter.
This formulation minimizes over features $\mathbf{z}_T, \mathbf{z}_R$ and two auxiliary parameters $\hat{\mathbf{T}}, \hat{\mathbf{R}}$. Based on Proximal Gradient Descent (PGD) \cite{beck2009fast,model_driven_rain_2020}, we propose an iterative algorithm to sequentially update $\mathbf{z}_T, \mathbf{z}_R, \hat{\mathbf{T}}, \hat{\mathbf{R}}$ with simple computations.

\textbf{Updating $\mathbf{z}_T$:} The sub-problem corresponding to $\mathbf{z}_T$ can be solved using quadratic approximation:
\begin{equation}
    \underset{\mathbf{z}_T}{\min} \frac{1}{2} \Vert \mathbf{z}_T - \left( \mathbf{z}_T^{(k)} - \eta_1 \nabla f(\mathbf{z}_T^{(k)})  \right) \Vert_F^2 +  \frac{\eta_1 \lambda_T}{\tau}  p_T(\mathbf{z}_T),
\end{equation}
where $\eta_1$ denotes the step-size for updating, the superscript $(k)$ denotes the results from the $k$-th iteration, and
$f(\mathbf{z}_T) = \frac{1}{2}\Vert \hat{\mathbf{T}} - \mathbf{D}_T \otimes \mathbf{z}_T \vert_F^2$. Therefore, its solution can be expressed as:
\begin{equation}
    \mathbf{z}_T^{(k+1)} = \text{prox}_{\eta_1 \lambda_T / \tau} \left( \mathbf{z}_T^{(k)} - \eta_1 \nabla f(\mathbf{z}_T^{(k)}) \right),
    \label{eq:solveZT}
\end{equation}
where $\text{prox}_{\eta_1 \lambda_T / \tau}(\cdot)$ is the proximal operator corresponding to the prior term $p_T(\cdot)$, $\nabla f(\mathbf{z}_T^{(k)}) = - \mathbf{D}_T^{(k)} \otimes^{T} ( \hat{\mathbf{T}} - \mathbf{D}_T^{(k)} \otimes \mathbf{z}_T^{(k)} )$, and $\otimes^{T}$ denotes the transposed convolution\footnote{The operation $\otimes^{T}$ can be implemented using the function “torch.nn.ConvTransposed2d” in PyTorch.}. 

\textbf{Updating $\mathbf{z}_R$:} The updating rule of $\mathbf{z}_R$ is similar to that of $\mathbf{z}_T$ and can be expressed as:
\begin{equation}
    \mathbf{z}_R^{(k+1)} = \text{prox}_{\eta_2 \lambda_R / \tau} \left( \mathbf{z}_R^{(k)} - \eta_2 \nabla h(\mathbf{z}_R^{(k)}) \right),
    \label{eq:solveZR}
\end{equation}
where $\eta_2$ denotes the step-size for updating, $\text{prox}_{\eta_2 \lambda_R / \tau}(\cdot)$ is the proximal operator corresponding to the prior term $p_R(\cdot)$, $\nabla h(\mathbf{z}_R^{(k)}) = - \mathbf{D}_R^{(k)} \otimes^{T} ( \hat{\mathbf{R}} - \hat{\mathbf{T}} - \mathbf{D}_R^{(k)} \otimes \mathbf{z}_R^{(k)} )$.

\textbf{Updating $\hat{\mathbf{T}}$:} The sub-problem with respect to $\hat{\mathbf{T}}$ can be expressed as:
\begin{equation}
    \begin{aligned}
        \underset{\hat{\mathbf{T}}}{\min} & \frac{1}{2} \Vert \mathbf{I} - \hat{\mathbf{T}} - \hat{\mathbf{R}} \Vert_F^2  +\frac{\tau}{2} \Vert \hat{\mathbf{T}} - \mathbf{D}_T \otimes \mathbf{z}_T \Vert_F^2 + \kappa \sum_{m=1}^{M} \Vert ( \mathbf{W}_m \otimes \hat{\mathbf{T}} ) \odot ( \mathbf{W}_m \otimes \hat{\mathbf{R}} ) \Vert_1.
    \end{aligned}
    \label{eq:Tsub}
\end{equation}

The quadratic approximation of Eq. (\ref{eq:Tsub}) can similarly be expressed as:
\begin{equation}
    \begin{aligned}
        \underset{\hat{\mathbf{T}}}{\min} & \frac{1}{2} \Vert \hat{\mathbf{T}} - ( \hat{\mathbf{T}}^{(k)} - \eta_3 \nabla u(\hat{\mathbf{T}}^{(k)})  ) \Vert_F^2  + \kappa \sum_{m=1}^{M} \Vert ( \mathbf{W}_m \otimes \hat{\mathbf{R}}^{(k)} ) \odot ( \mathbf{W}_m \otimes \hat{\mathbf{T}} ) \Vert_1,
    \end{aligned}
    \label{eq:Tquad}
\end{equation}
where $u(\hat{\mathbf{T}})=\frac{1}{2} \Vert \mathbf{I} - \hat{\mathbf{T}} - \hat{\mathbf{R}}^{(k)} \Vert_F^2  +\frac{\tau}{2} \Vert \hat{\mathbf{T}} - \mathbf{D}_T \otimes \mathbf{z}_T^{(k+1)} \Vert_F^2$. Therefore $\nabla u(\hat{\mathbf{T}}) = -(\mathbf{I} - \hat{\mathbf{R}}^{(k)} - \hat{\mathbf{T}} ) +   \tau ( \hat{\mathbf{T}} - \mathbf{D}_T \otimes \mathbf{z}_T^{(k+1)} )$.

When optimizing with respect to $\hat{\mathbf{T}}$, the estimated reflection image $\hat{\mathbf{R}}$ is assumed to be fixed. Therefore, the transform coefficients of the reflection image $\mathbf{W}_m \otimes \hat{\mathbf{R}}$ in the proposed transform-based exclusion loss can be treated as an element-wise regularization parameter for the transform coefficients $\mathbf{W}_m \otimes \hat{\mathbf{T}}$ of the transmission image. 
Consequently, the solution to Eq. (\ref{eq:Tsub}) can be expressed in terms of the proximal operator for the proposed transform-based exclusion loss:
\begin{equation}
    \hat{\mathbf{T}}^{(k+1)} = \sum_{m=1}^{M} \mathbf{W}_m^{\dagger} \otimes \mathcal{S}_{\kappa |\mathbf{W}_m \otimes \hat{\mathbf{R}}^{(k)}|} ( \mathbf{W}_m \otimes \phi(\hat{\mathbf{T}}^{(k)}) ),
    \label{eq:solveT}
\end{equation}
where $\phi(\hat{\mathbf{T}}^{(k)}) = \hat{\mathbf{T}}^{(k)} - \eta_3 \nabla u(\hat{\mathbf{T}}^{(k)})$ and $\mathbf{W}_m^{\dagger}$ denotes the inverse filter of $\mathbf{W}_m$.

The proximal operator is the soft-thresholding operator performed on the transform coefficients of $\phi(\hat{\mathbf{T}})^{(k)}$. The soft-thresholds ${\kappa |\mathbf{W}_m \otimes \hat{\mathbf{R}}^{(k)}|}$ is position dependent and based on the transform coefficients of the estimated reflection image $\hat{\mathbf{R}}^{(k)}$. After soft-thresholding, the updated transmission image is reconstructed using inverse transform with the soft-thresholded transform coefficients.

\textbf{Updating $\hat{\mathbf{R}}$:} Similar to the updating rule for $\hat{\mathbf{T}}$, we can express the solution to the sub-problem corresponding to $\hat{\mathbf{R}}$ as follows:
\begin{equation}
    \hat{\mathbf{R}}^{(k+1)} = \sum_{m=1}^{M} \mathbf{W}_m^{\dagger} \otimes \mathcal{S}_{\kappa |\mathbf{W}_m \otimes \hat{\mathbf{T}}^{(k+1)}|} ( \mathbf{W}_m \otimes \psi(\hat{\mathbf{R}}^{(k)})),
    \label{eq:solveR}
\end{equation}
where $\psi(\hat{\mathbf{R}}^{(k)}) = \hat{\mathbf{R}}^{(k)} - \eta_4 \nabla v(\hat{\mathbf{R}}^{(k)})$ and 
$\nabla v(\hat{\mathbf{R}}) = -(\mathbf{I} - \hat{\mathbf{R}} - \hat{\mathbf{T}}) + \tau (\hat{\mathbf{R}} - \mathbf{D}_R \otimes \mathbf{z}_R^{(k+1)})$.

\subsection{Deep Unfolded Reflection Removal Network (DURRNet)}
\label{sec:network}

\begin{figure}[t]
\begin{centering}
\center\includegraphics[width=0.95\columnwidth]{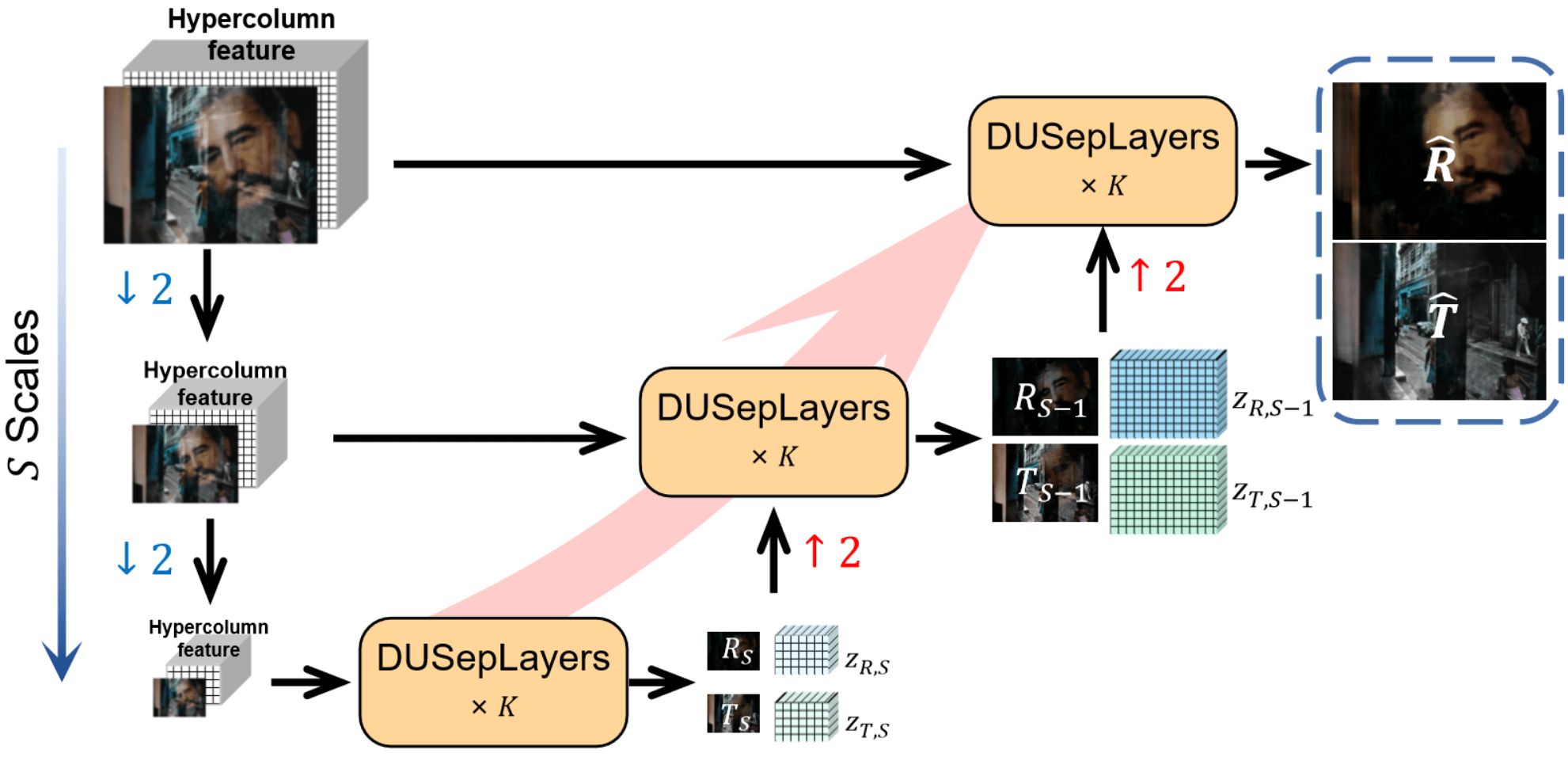}
\par\end{centering}
\caption{The proposed Deep Unfolded Reflection Removal Network (DURRNet). It consists of $S$ scales of DURRLayers to gradually estimate the transmission and the reflection images from low-resolution scales to the resolution of the input image. At each scale, there are $K$ stages of DURRLayers.
$\downarrow 2$ and $\uparrow 2$ denotes bilinear interpolation by a factor of 0.5 and 2, respectively.
}
\label{fig:DUSepNet}
\end{figure}

In this section, by using the unfolding technique, we construct a model-driven multi-scale Deep Unfolded Reflection Removal Network (DURRNet) with multiple Deep Unfolded Reflection Removal Layers (DURRLayers). Each DURRLayer unrolls an iteration of the proposed iterative algorithm for single image reflection removal. 

\textbf{Overall Architecture:} 
As shown in Fig. \ref{fig:DUSepNet}, the proposed DURRNet is designed in a multi-resolution fashion. There are $S$ scales of DURRLayers to effectively exploit information at different scales for separating the input image into a transmission image and a reflection image. Each scale consists of $K$ DURRLayers. At the lowest scale, the initial transmission image $\mathbf{T}_S$, reflection image $\mathbf{R}_S$ and features $\mathbf{z}_{T,S},\mathbf{z}_{R,S}$ are initialized based on the down-sampled input image and the the hyper-column feature \cite{perceptual_loss_2018,hariharan2015hypercolumns} of the input image using bilinear interpolation by a factor $2^{S-1}$, respectively. At an upper scale, the transmission and reflection images are initialized based on the $2$ times up-sampled version estimated from its lower scale, and the features are initialized based on the down-sampled hyper-column feature of the input image and the up-sampled feature estimated from the lower scale. The multi-scale architecture performs image separation in a coarse-to-fine manner and can therefore effectively combine information from different scales.

\textbf{DURRLayer:} 
Fig. \ref{fig:DUSepLayer} shows the network structure for the proposed Deep Unfolded Reflection Removal Layer (DURRLayer) which corresponds to one iteration of the proposed iterative algorithm.
The model-inspired DURRLayer enables that the estimated transmission and reflection image can well reconstruct the input image and the prior information can be properly imposed.
For each image layer, a proximal network ProxNet is used to impose the prior for the feature, and a proximal network based on invertible network ProxInvNet is proposed to impose the exclusion prior for each estimated image.

\textbf{ProxNet:}
Similar to \cite{model_driven_rain_2020}, the proximal operators for $\mathbf{z}_T$ and $\mathbf{z}_R$ in Eq. (\ref{eq:solveZT}) and (\ref{eq:solveZR}) are represented by two deep convolutional networks $\text{ProxNet}_{\mathbf{\theta}_{\mathbf{z}_T}}(\cdot)$ and $\text{ProxNet}_{\mathbf{\theta}_{\mathbf{z}_R}}(\cdot)$ 
whose parameters are learned from the training dataset to well capture the prior information. 
The updating rule for $\mathbf{z}_T$ and $\mathbf{z}_R$ can be therefore expressed as:
\begin{equation}
    \begin{cases}
        \nabla f(\mathbf{z}_T^{(k)}) = - \mathbf{K}_T^{(k)} \otimes^{T} \left( \hat{\mathbf{T}}^{(k)} - \mathbf{D}_T^{(k)} \otimes \mathbf{z}_T^{(k)} \right), \\
        \mathbf{z}_T^{(k+1)} = \text{ProxNet}_{\mathbf{\theta}_{\mathbf{z}_T}} \left( \mathbf{z}_T^{(k)} - \nabla f(\mathbf{z}_T^{(k)}) \right), \\
    \end{cases}
\end{equation}
\begin{equation}
    \begin{cases}
        \nabla h(\mathbf{z}_R^{(k)}) = - \mathbf{K}_R^{(k)} \otimes^{T} \left( \hat{\mathbf{R}}^{(k)} - \mathbf{D}_R^{(k)} \otimes \mathbf{z}_R^{(k)} \right), \\
        \mathbf{z}_R^{(k+1)} = \text{ProxNet}_{\mathbf{\theta}_{\mathbf{z}_R}} \left( \mathbf{z}_R^{(k)} - \nabla f(\mathbf{z}_R^{(k)}) \right),
    \end{cases}
\end{equation}
where the convolutional dictionaries $\mathbf{D}_T^{(k)}$, $\mathbf{D}_R^{(k)}$, $\mathbf{K}_T^{(k)}$, and $\mathbf{K}_R^{(k)}$ and the parameters of the proximal networks $\mathbf{\theta}_{\mathbf{z}_T}$ and $\mathbf{\theta}_{\mathbf{z}_R}$ are learnable parameters.

\begin{figure}[t]
\begin{centering}
\center\includegraphics[width=0.8\columnwidth]{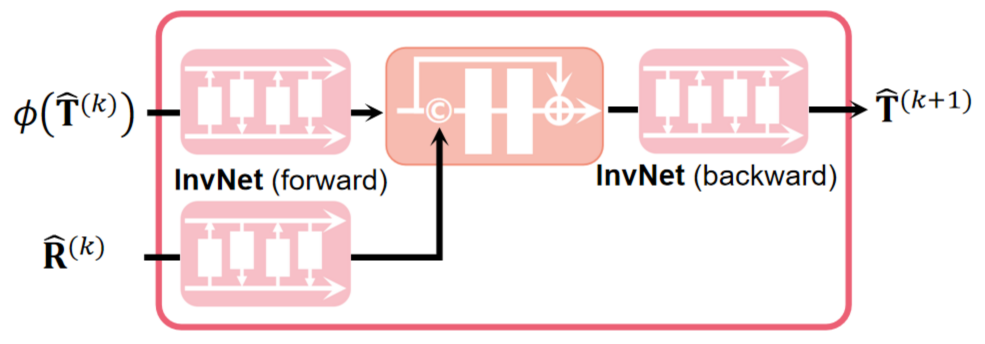}
\par\end{centering}
\caption{The proposed Proximal Invertible Network $\text{ProxInvNete}_{\mathbf{\theta}_{\mathbf{T}}}(\cdot, \cdot)$. The invertible network (invNet) serves as an invertible transform to transform images to coefficient domain using its forward pass then transform the coefficients back to image domain using its backward pass. 
}
\label{fig:proxInvNet}
\end{figure}

\textbf{ProxInvNet:}
For the proximal operators for the transform-based exclusion term, a direct option is to apply wavelet transform to extract edge information, use soft-thresholding operator to suppress common content and then reconstruct the image using the inverse wavelet transform. However, the fixed transform may not be sufficiently flexible to handle complex reflections.
Inspired by the invertible networks as a learnable invertible transform \cite{huang2021linn,huang2021winnet}, we propose to use the invertible networks to construct a learnable proximal operator $\text{ProxInvNete}_{\mathbf{\theta}}(\cdot)$ for imposing the exclusion condition. 
The forward pass of the invertible networks serves as the forward transform, and the backward pass of the invertible networks then serves as the corresponding inverse transform. 
The updating rule for $\hat{\mathbf{T}}$ and $\hat{\mathbf{R}}$ can be expressed as:
\begin{equation}
    \begin{cases}
        \mathcal{E}_{\mathbf{T}}^{(k+1)} = \hat{\mathbf{T}}^{(k)} - \mathbf{D}_T^{(k)} \otimes \mathbf{z}_T^{(k+1)} ,\\
        \phi(\hat{\mathbf{T}}^{(k)}) = \hat{\mathbf{T}}^{(k)} + \eta_T \left((\mathbf{I} - \hat{\mathbf{R}}^{(k)} - \hat{\mathbf{T}}^{(k)} ) - \tau_T \mathcal{E}_{\mathbf{T}}^{(k+1)}
        \right), \\
        \hat{\mathbf{T}}^{(k+1)} = \text{ProxInvNet}_{\mathbf{\theta}_{\mathbf{T}}} \left( \phi(\hat{\mathbf{T}}^{(k)}), \hat{\mathbf{R}^{(k)}} \right), \\
    \end{cases}
\end{equation}
\begin{equation}
    \begin{cases}
        \mathcal{E}_{\mathbf{R}}^{(k+1)} = \hat{\mathbf{R}} - \mathbf{D}_R^{(k)} \otimes \mathbf{z}_R^{(k+1)},\\
        \psi(\hat{\mathbf{R}}^{(k)}) = \hat{\mathbf{R}}^{(k)} + \eta_R \left((\mathbf{I} - \hat{\mathbf{R}}^{(k)} - \hat{\mathbf{T}}^{(k+1)}) - \tau_R \mathcal{E}_{\mathbf{R}}^{(k+1)} \right),\\
        \hat{\mathbf{R}}^{(k+1)} = \text{ProxInvNet}_{\mathbf{\theta}_{\mathbf{R}}} \left( \psi(\hat{\mathbf{R}}^{(k)}), \hat{\mathbf{T}}^{(k+1)} \right),
    \end{cases}
\end{equation}
where the convolutional dictionaries $\mathbf{D}_T^{(k)}$ and $\mathbf{D}_R^{(k)}$, and step size parameter $\eta_T$, $\eta_R$, $\tau_T$, and $\tau_R$ and the parameters of the proximal invertible networks $\mathbf{\theta}_{\mathbf{T}}$ and $\mathbf{\theta}_{\mathbf{R}}$ are learnable parameters.

Fig. \ref{fig:proxInvNet} shows the diagram for $\text{ProxInvNete}_{\mathbf{\theta}_{\mathbf{T}}}(\cdot, \cdot)$.
The forward pass of the invertible networks is applied as the forward transform to extract features from $\phi(\hat{\mathbf{T}}^{(k)})$ and $\hat{\mathbf{R}^{(k)}}$.
In the Threshold Network (ThreNet), the feature of $\hat{\mathbf{R}}$ will be concatenated with that of $\phi(\hat{\mathbf{T}}^{(k)})$ and then they pass through a convolutional network with residual blocks to generate corrections for the feature of $\hat{\mathbf{T}}$. The updated feature of $\hat{\mathbf{T}}$ will then be converted back to image domain using the backward pass of the invertible networks. Similar operations can be performed when updating $\hat{\mathbf{R}}$. The forward and backward pass of the invertible networks are constructed by the same set of $P$ pairs of prediction and updater networks (PUNet), for details please refer to \cite{huang2021winnet}.




\begin{figure}[t]
    \centering
    \includegraphics[width=1\columnwidth]{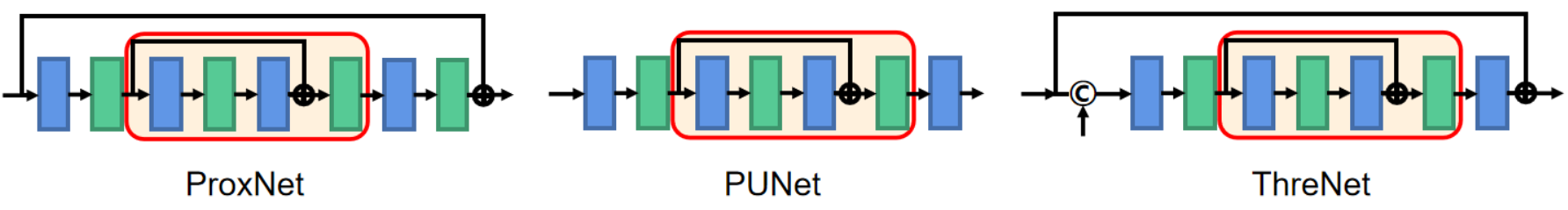}
    \caption{The network architectures for ProxNet, PUNet and ThreNet used to construct the proposed DURRLayer. The blue and green blocks represent convolutional layers and ReLU activation layers, respectively. The yellow blocks represent a residual block.}
    \label{fig:networks}
\end{figure}

\subsection{Training Details}
\label{sec:training}
Apart from the proposed exclusion loss we introduced in Section 3.1, we adopt the reconstruction loss and the perceptual loss~\cite{perceptual_loss_2018} for training:
\begin{equation}
    \mathcal{L}= \mathcal{L}_{ {r}} + \lambda_{e} \mathcal{L}_{{e}} + \lambda_{p} \mathcal{L}_{ {p}},
\end{equation}
where $\lambda_{e}=0.01$ and $\lambda_{p}=0.01$ are regularization parameters. The reconstruction loss $\mathcal{L}_{\text {r}}$ is applied to the estimated transmission image $\widehat{\mathbf{T}}$ and reflection image $\widehat{\mathbf{R}}$ as well as the reconstructed image based on the final features:
\begin{equation}
\begin{aligned}
    \mathcal{L}_{{r}}= & \|\mathbf{T}-\widehat{\mathbf{T}}\|_{2}^{2} + \|\mathbf{T}-\mathbf{D}_T \otimes \mathbf{z}_T\|_{2}^{2}  + \|\mathbf{R}-\hat{\mathbf{R}}\|_{2}^{2}+\|\mathbf{R}-\mathbf{D}_R \otimes \mathbf{z}_R\|_{2}^{2}.
\end{aligned}
\end{equation}
Perceptual loss~\cite{perceptual_loss_2018} is used to regularize the estimated images with high perceptual quality by minimizing the $l_1$ difference between the VGG features of the estimated and the ground-truth images:
\begin{equation}
    \mathcal{L}_{ {p}}=\|\tau(\mathbf{T})-\tau(\widehat{\mathbf{T}})\|_{1}+\|\tau(\mathbf{R})-\tau(\widehat{\mathbf{R}})\|_{1},
\end{equation}
where $\tau(\cdot)$ denotes the features of the VGG-19 model pretrained on ImageNet dataset.



\section{Experimental Results}
\label{sec:results}


\subsection{Implementation Details}

The proposed method is implemented with Pytorch, and the models are optimized with Adam optimizer with initial learning rate $10^{-4}$ which are decayed at epoch 10, 15, and 20 with learning rate decay 0.5. The total number of epochs is 25. The early stop strategy is used.
The experiments were performed on a computer with a RTX 3090 Ti GPU.

The number of scales $S$ in DURRNet is set to 4 and the number of DURRLayer stages in each scale is set to 2. The number of feature channels is set to 64. The forward pass and backward pass of the invertible networks consists of $P=2$ pairs of PUNets. 
The network architectures of ProxNet, PUNet, ThreNet used to construct DURRLayer are illustrated in Fig. \ref{fig:networks}. All the networks are constructed using convolutional layers, ReLU layers and Residual blocks.

\begin{table}[t]
\center
\caption{Quantitative comparisons on Real20 testing dataset~\cite{perceptual_loss_2018} of different methods. (The best scores are in bold.)}
\begin{tabular}{C{1.6cm}|C{1.6cm}C{1.6cm}C{1.6cm}C{1.6cm}C{1.6cm}C{1.6cm}}
\toprule
Metrics  & CEILNet & Zhang \textit{et al.} & BDN & IBCLN & YTMT & DURRNet   \\ \hline \hline

PSNR & 18.45 & 22.55 & 18.41 & 21.86 & 23.26 & \textbf{23.61} \\
SSIM & 0.690 & 0.788 & 0.726 & 0.762 & \textbf{0.806} & 0.804 \\ \hline




\bottomrule
\end{tabular}

\label{tab:sota}
\end{table}

\begin{table}[t]
\center
\caption{Quantitative comparisons on Nature testing dataset~\cite{cascaded_refine_2020} of different methods. (The best scores are in bold.)}
\begin{tabular}{C{1.6cm}|C{1.6cm}C{1.6cm}C{1.6cm}C{1.6cm}C{1.6cm}C{1.6cm}}
\toprule
Metrics  & CEILNet-F & Zhang \textit{et al.} & BDN-F & IBCLN & YTMT & DURRNet   \\ \hline \hline

PSNR & 19.33 & 19.56 & 18.92 & 23.57 & {23.85} & \textbf{24.29} \\ 
SSIM & 0.745 & 0.736 & 0.737 & 0.783 & \textbf{0.810} & 0.806 \\ \hline 
\bottomrule
\end{tabular}

\label{tab:sota_nature}
\end{table}

\begin{figure}[t]
    \centering
    \subfigure[{Zhang \textit{et al.}}]{
    \begin{minipage}[b]{0.17\textwidth}
    \includegraphics[width=1.0\linewidth]{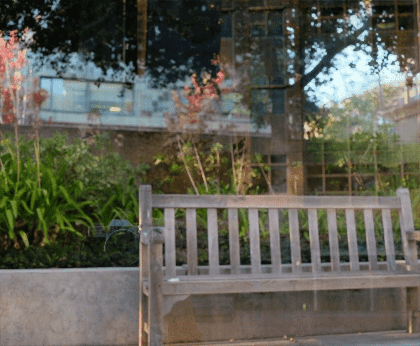}\vspace{0.01pt}
    \includegraphics[width=1.0\linewidth]{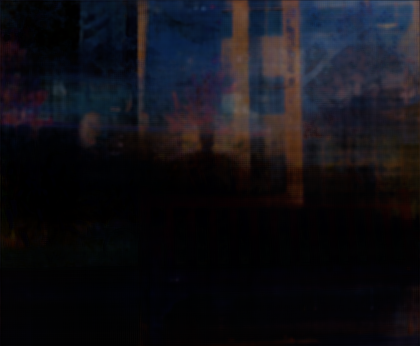}\vspace{2pt}
    \includegraphics[width=1.0\linewidth]{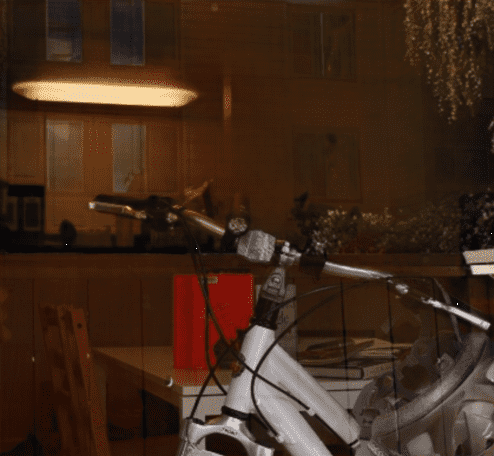}\vspace{0.01pt}
    \includegraphics[width=1.0\linewidth]{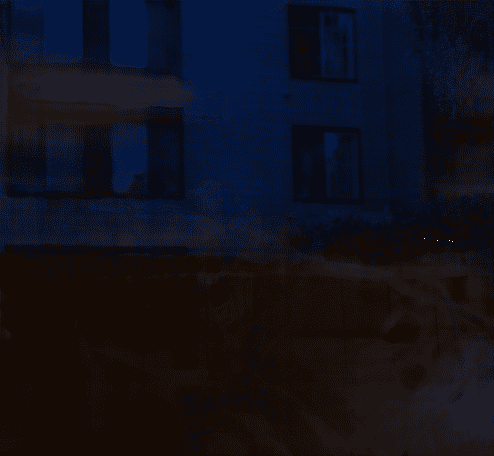}\vspace{2pt}
    \includegraphics[width=1.0\linewidth]{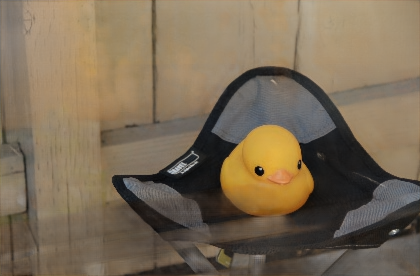}\vspace{0.01pt}
    \includegraphics[width=1.0\linewidth]{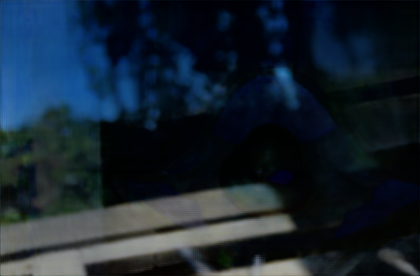}\vspace{2pt}
    \end{minipage}
    }
    \subfigure[BDN]{
    \begin{minipage}[b]{0.17\textwidth}
    \includegraphics[width=1.0\linewidth]{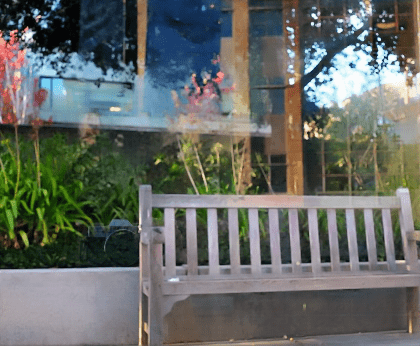}\vspace{0.01pt}
    \includegraphics[width=1.0\linewidth]{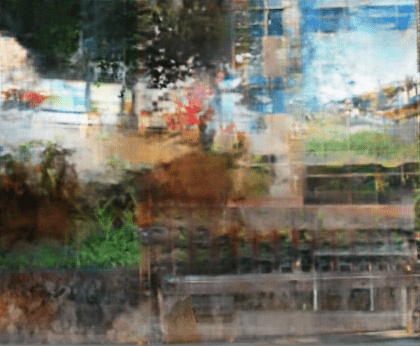}\vspace{2pt}
    \includegraphics[width=1.0\linewidth]{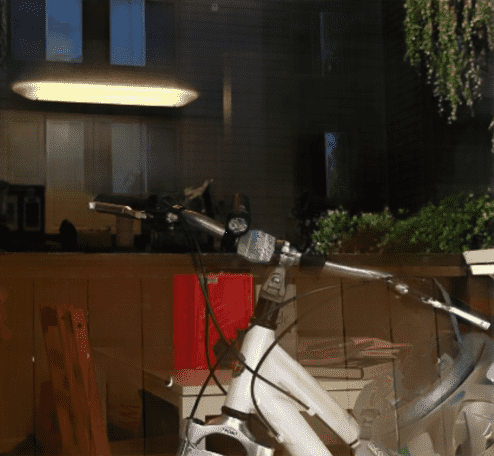}\vspace{0.01pt}
    \includegraphics[width=1.0\linewidth]{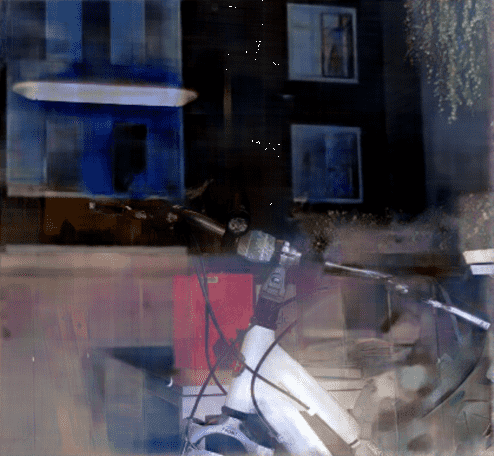}\vspace{2pt}
    \includegraphics[width=1.0\linewidth]{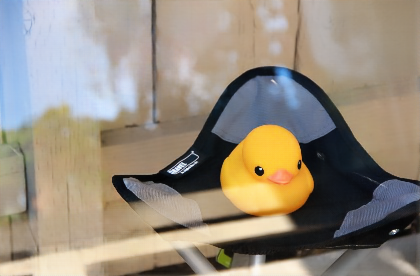}\vspace{0.01pt}
    \includegraphics[width=1.0\linewidth]{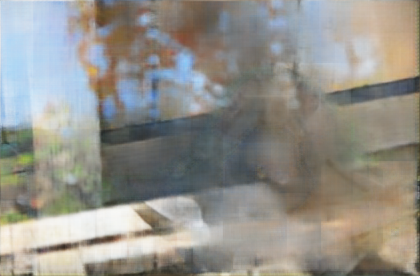}\vspace{2pt}
    \end{minipage}
    }
    \subfigure[{IBCLN}]{
    \begin{minipage}[b]{0.17\textwidth}
    \includegraphics[width=1.0\linewidth]{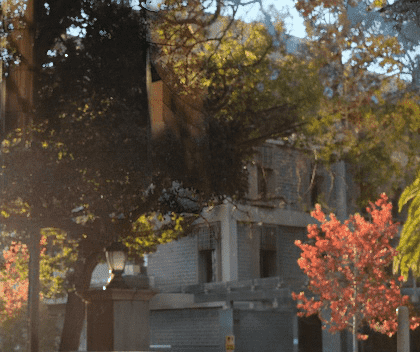}\vspace{0.01pt}
    \includegraphics[width=1.0\linewidth]{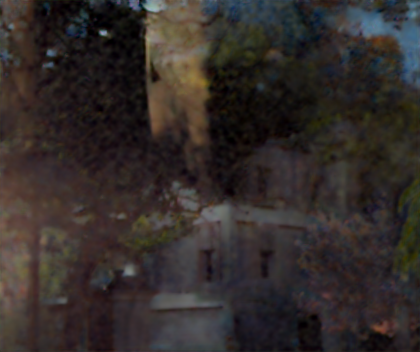}\vspace{2pt}
    \includegraphics[width=1.0\linewidth]{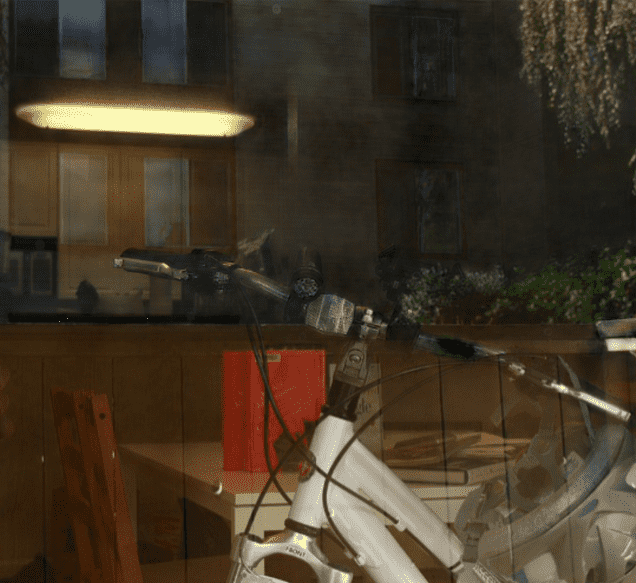}\vspace{0.01pt}
    \includegraphics[width=1.0\linewidth]{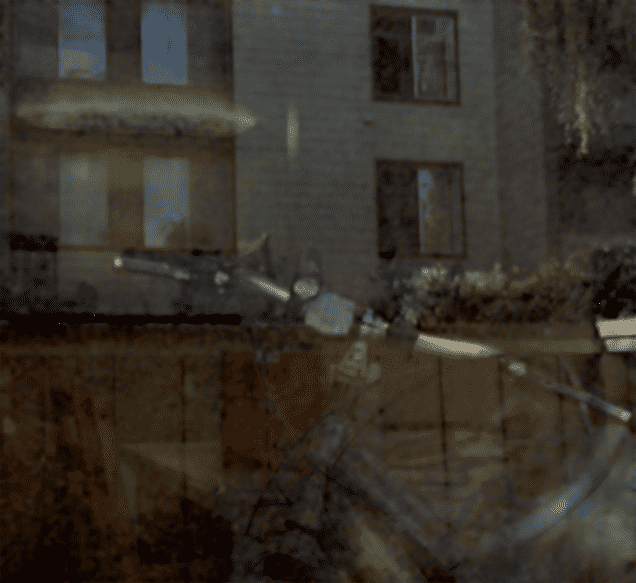}\vspace{2pt}
    \includegraphics[width=1.0\linewidth]{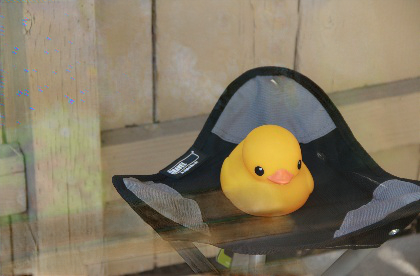}\vspace{0.01pt}
    \includegraphics[width=1.0\linewidth]{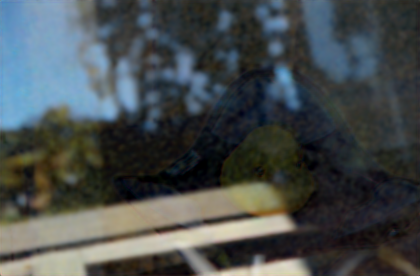}\vspace{2pt}
    \end{minipage}
    }
    \subfigure[DURRNet]{
    \begin{minipage}[b]{0.17\textwidth}
    \includegraphics[width=1.0\linewidth]{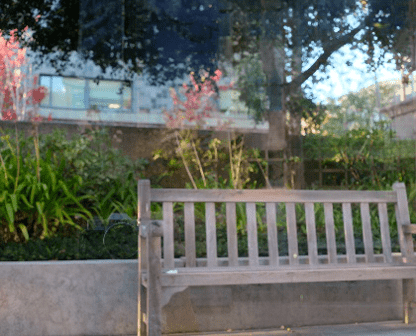}\vspace{0.01pt}
    \includegraphics[width=1.0\linewidth]{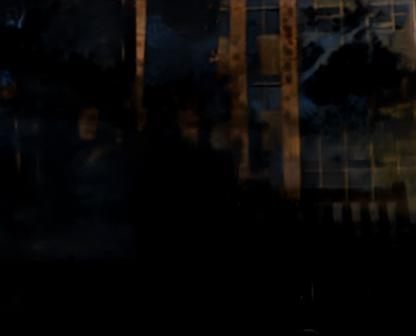}\vspace{2pt}
    \includegraphics[width=1.0\linewidth]{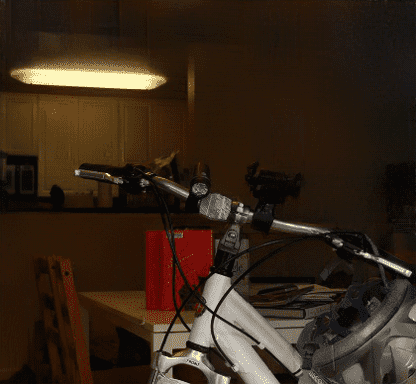}\vspace{0.01pt}
    \includegraphics[width=1.0\linewidth]{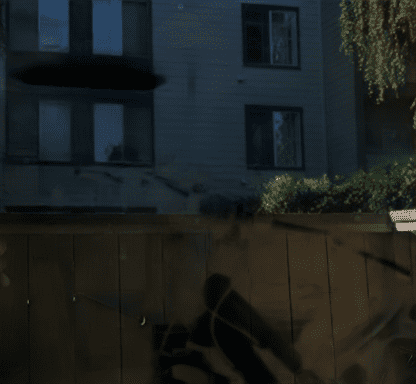}\vspace{2pt}
    \includegraphics[width=1.0\linewidth]{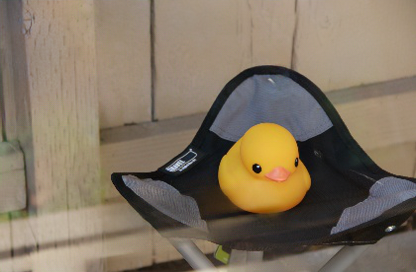}\vspace{0.01pt}
    \includegraphics[width=1.0\linewidth]{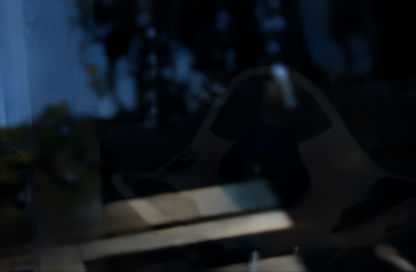}\vspace{2pt}
    \end{minipage}
    }
    \subfigure[GT]{
    \begin{minipage}[b]{0.17\textwidth}
    \includegraphics[width=1.0\linewidth]{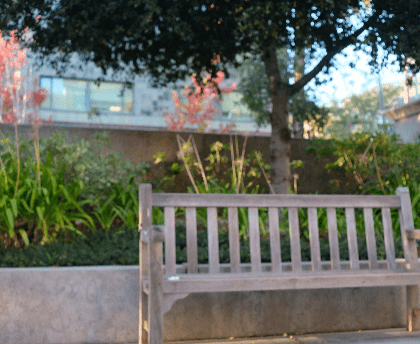}\vspace{0.01pt}
    \includegraphics[width=1.0\linewidth]{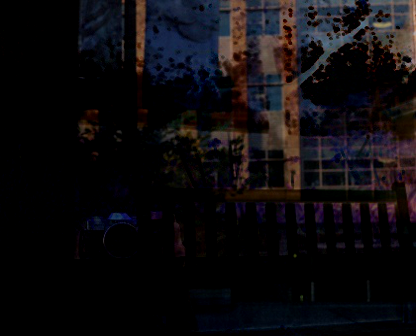}\vspace{2pt}
    \includegraphics[width=1.0\linewidth]{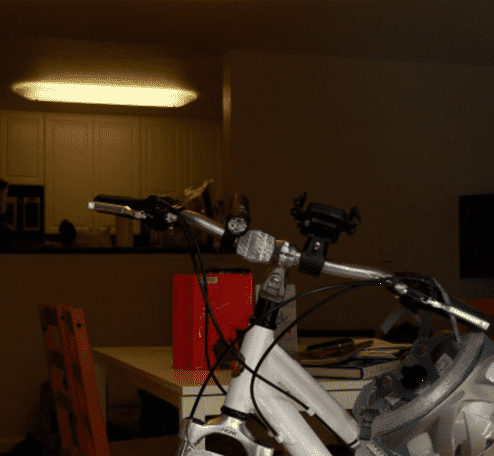}\vspace{0.01pt}
    \includegraphics[width=1.0\linewidth]{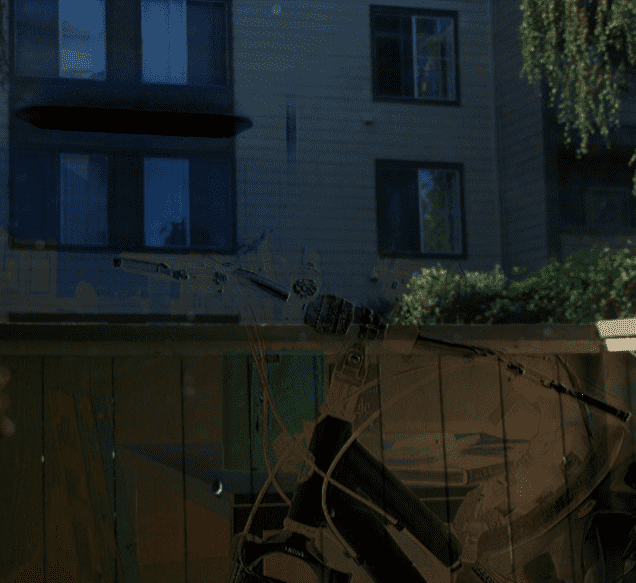}\vspace{2pt}
    \includegraphics[width=1.0\linewidth]{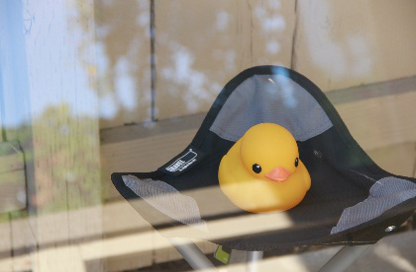}\vspace{0.01pt}
    \includegraphics[width=1.0\linewidth]{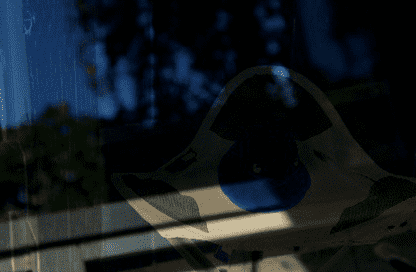}\vspace{2pt}
    \end{minipage}
    }
    
   \caption{Visual comparisons on the estimated transmission image (row 1, 3 and 5) and the estimated reflection image (row 2, 4 and 6) of different single image reflection methods on Real20 dataset~\cite{perceptual_loss_2018}. The last column shows the ground-truth transmission image and reflection image for the reference.}
    \label{fig:real20}
\end{figure}

\subsection{Comparison with State-of-the-arts Methods}

In this section, we quantitatively and visually compare our DURRNet with other single image reflection removal methods including CEILNet method~\cite{generic_smooth_2017}, Zhang \textit{et al.}'s method~\cite{perceptual_loss_2018}, BDN method~\cite{yang2018seeing}, IBCLN method~\cite{cascaded_refine_2020} and YTMT method~\cite{hu2021trash}.

Table \ref{tab:sota} shows the quantitative evaluation results of different single image reflection removal methods evaluated on Real20 dataset~\cite{perceptual_loss_2018}. 
The training datasets consist of synthetically generated reflection-contaminated images using 7643 image pairs from PASCAL VOC dataset by following the settings in CEILNet~\cite{generic_smooth_2017} and 90 pairs of real images from ~\cite{perceptual_loss_2018}. 
The testing datasets contain 20 images from Real20~\cite{perceptual_loss_2018}.
From Table \ref{tab:sota}, we can see that on Real20 dataset the proposed DURRNet achieves significantly better PSNR compared to other methods and achieves a similar SSIM results as YTMT method.

Table \ref{tab:sota_nature} shows the quantitative comparison results on Nature testing dataset~\cite{cascaded_refine_2020}. The comparison follows the settings in~\cite{cascaded_refine_2020}. Additional 200 training image pairs from the Nature training dataset~\cite{cascaded_refine_2020} were used for training and other models (with a suffix ``-F") were fine-tuned on the Nature training dataset for fair comparisons. We can see that the proposed DURRNet achieves the highest PSNR value and the second best SSIM value among all the methods.

For visual comparisons, Fig. \ref{fig:real20} shows the estimated transmission and reflection images by different methods on 3 exemplar images from Real20 dataset~\cite{perceptual_loss_2018}. 
This dataset is a challenging dataset since the input images contain different reflection patterns and the region of overlap is large.
From Fig. \ref{fig:real20}, we can see that the proposed DURRNet is able to recover natural looking transmission and reflection images. This could be due to the fact that the proposed deep unfolded network architecture takes the image formation model into consideration and prior information has been properly imposed into the network architecture. 
For comparison methods, Zhang \textit{et al.}'s method is able to well separate most reflections in the input, but may generate images with visible artifacts, BDN method did not successfully remove strong reflections and usually generates reflection images with too much transmission image content, and IBCLN method struggle to separate large overlapping reflections.

\begin{figure}[t]
    \centering
    \subfigure[{Input}]{
    \begin{minipage}[b]{0.17\textwidth}
    \includegraphics[width=1.0\linewidth]{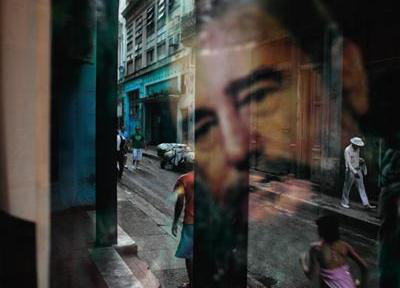}\vspace{0.01pt}
    \includegraphics[width=1.0\linewidth]{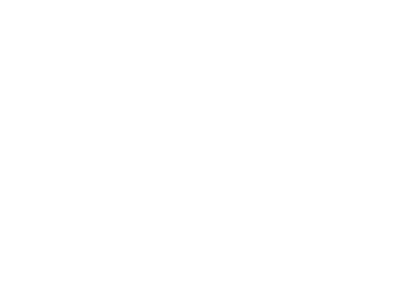}\vspace{2pt}
    \includegraphics[width=1.0\linewidth]{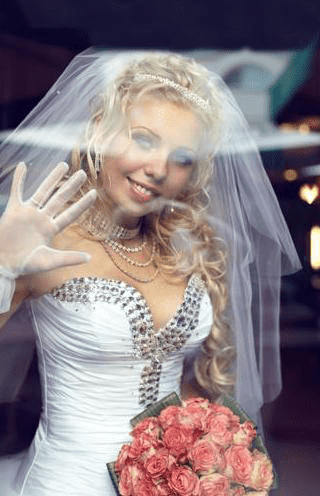}\vspace{0.01pt}
    \includegraphics[width=1.0\linewidth]{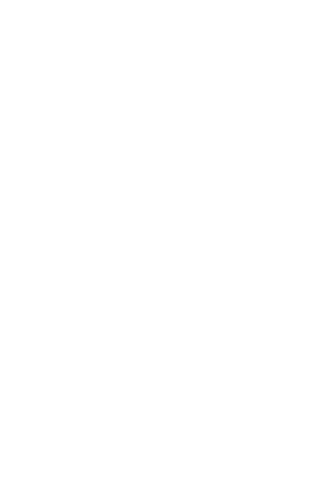}\vspace{2pt}
    \end{minipage}
    }
    \subfigure[Zhang \textit{et al.}]{
    \begin{minipage}[b]{0.17\textwidth}
    \includegraphics[width=1.0\linewidth]{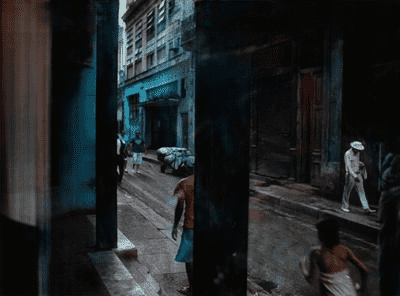}\vspace{0.01pt}
    \includegraphics[width=1.0\linewidth]{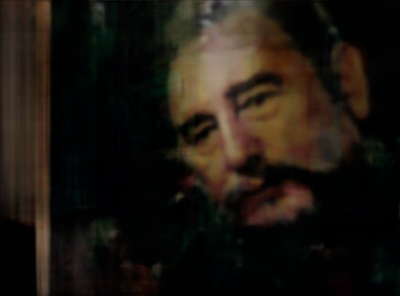}\vspace{2pt}
    \includegraphics[width=1.0\linewidth]{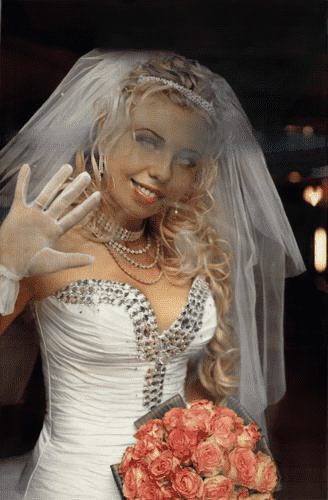}\vspace{1.2pt}
    \includegraphics[width=1.0\linewidth]{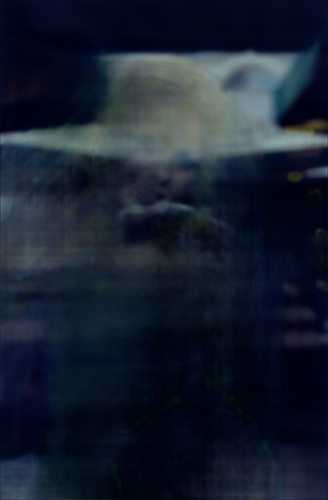}\vspace{2pt}
    \end{minipage}
    }
    \subfigure[{BDN}]{
    \begin{minipage}[b]{0.17\textwidth}
    \includegraphics[width=1.0\linewidth]{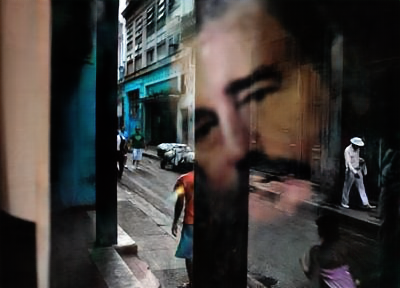}\vspace{0.01pt}
    \includegraphics[width=1.0\linewidth]{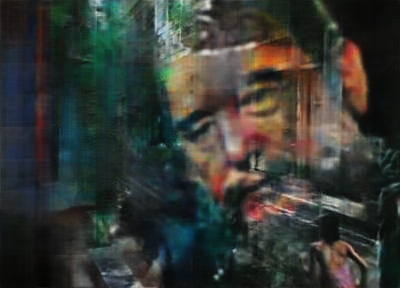}\vspace{2pt}
    \includegraphics[width=1.0\linewidth]{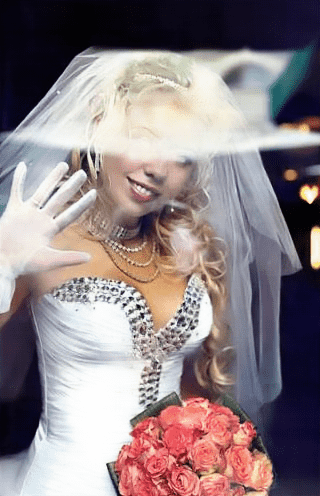}\vspace{0.01pt}
    \includegraphics[width=1.0\linewidth]{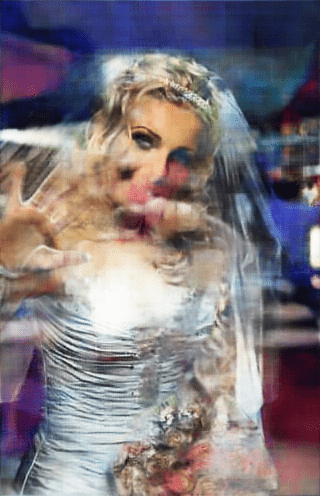}\vspace{2pt}
    \end{minipage}
    }
    \subfigure[IBCLN]{
    \begin{minipage}[b]{0.17\textwidth}
    \includegraphics[width=1.0\linewidth]{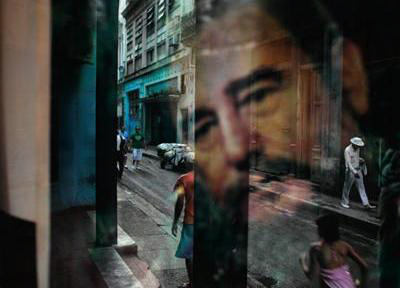}\vspace{0.01pt}
    \includegraphics[width=1.0\linewidth]{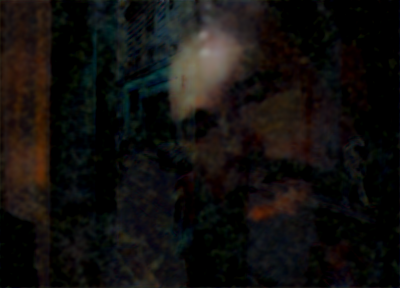}\vspace{2pt}
    \includegraphics[width=1.0\linewidth]{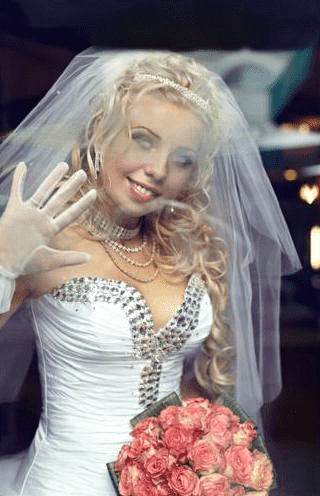}\vspace{0.01pt}
    \includegraphics[width=1.0\linewidth]{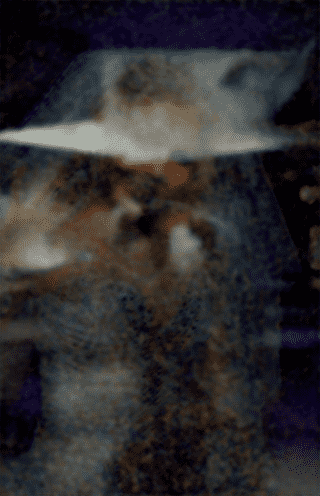}\vspace{2pt}
    \end{minipage}
    }
    \subfigure[DURRNet]{
    \begin{minipage}[b]{0.17\textwidth}
    \includegraphics[width=1.0\linewidth]{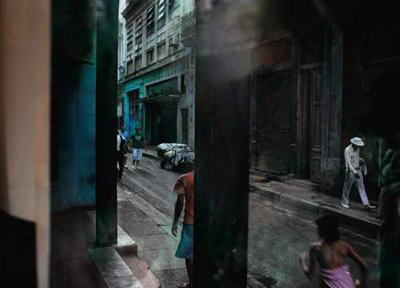}\vspace{0.01pt}
    \includegraphics[width=1.0\linewidth]{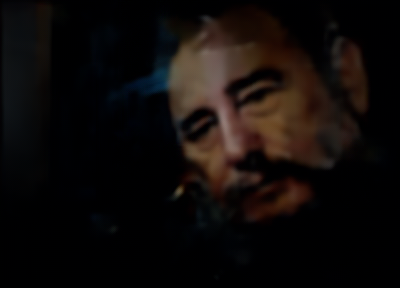}\vspace{2pt}
    \includegraphics[width=1.0\linewidth]{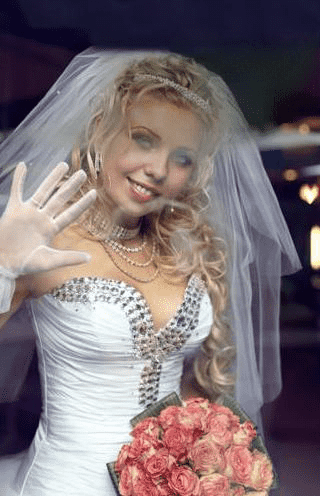}\vspace{0.01pt}
    \includegraphics[width=1.0\linewidth]{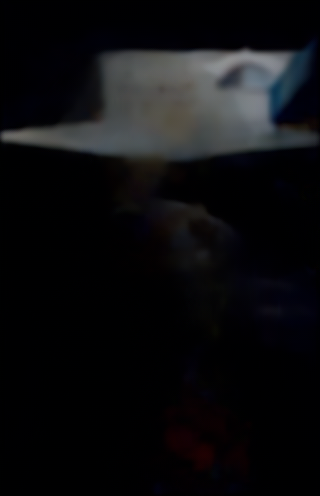}\vspace{2pt}
    \end{minipage}
    }
    \caption{Visual comparisons of different single image reflection methods on Real45 dataset~\cite{generic_smooth_2017}. Row 1 and 3 show the estimated transmission images, row 2 and 4 show the estimated reflection image. }
    \label{fig:real45}
\end{figure}

Fig. \ref{fig:real45} further shows the visual comparisons on Real45 dataset~\cite{generic_smooth_2017} which does not contain ground-truth images for reference. We can see that the proposed DURRNet is able to properly separate the reflection image content from the input reflection-contaminated image and the separated reflection images contain little information from the transmission image.

\subsection{Ablation Studies}

\noindent \textbf{The effectiveness of ProxNet/ProxInvNet:} In the proposed DURRLayer, the main network components are the ProxNet and ProxInvNet which is used to impose natural image prior and exclusion prior, respectively. 
To understand their functionalities, we perform ablation studies on these network components.

\begin{table}[]
\center
\caption{Quantitative performance of the proposed DURRNet with different variations. The performance of different models are evaluated on Real20 dataset~\cite{perceptual_loss_2018}.}
\begin{tabular}{l|C{2cm}|C{2cm}|c|c|c}
\toprule
Settings & DURRNet & w/o ProxNet  & w/o ProxInvNet     & $(S,K)=(1,8)$ & $(S,K)=(2,4)$
\\ \hline \hline
PSNR    & 23.61     & 22.63     & 22.61     & 22.47   & 22.74     \\ \hline
SSIM    & 0.803     & 0.787     & 0.788     & 0.787   & 0.794     \\ \hline
\bottomrule

\end{tabular}
\label{tab:ablation}
\end{table}

\begin{figure}
    \centering
    \subfigure[Input]{
    \begin{minipage}[b]{ 0.23\textwidth}
    \includegraphics[width=1.0\linewidth]{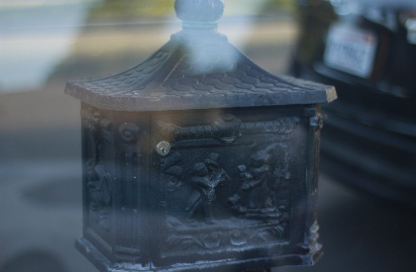}\vspace{0.1pt}
    \includegraphics[width=1.0\linewidth]{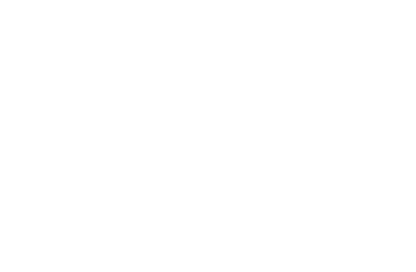}\vspace{2pt}
    \end{minipage}
    }
    \subfigure[w/o ProxNet]{
    \begin{minipage}[b]{ 0.23\textwidth}
    \includegraphics[width=1.0\linewidth]{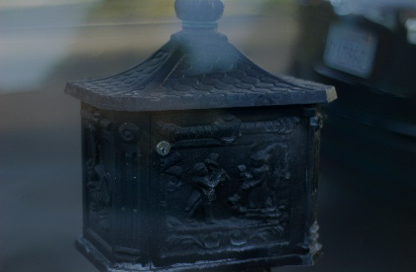}\vspace{0.1pt}
    \includegraphics[width=1.0\linewidth]{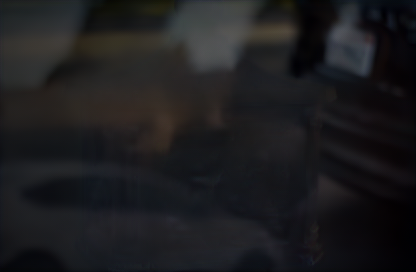}\vspace{2pt}
    \end{minipage}
    }
    \subfigure[w/o {ProxInvNet}]{
    \begin{minipage}[b]{ 0.23\textwidth}
    \includegraphics[width=1.0\linewidth]{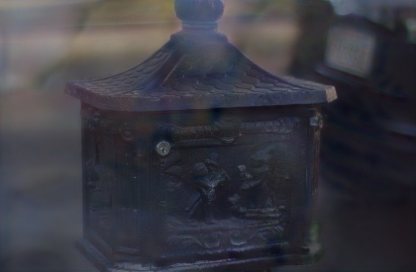}\vspace{0.1pt}
    \includegraphics[width=1.0\linewidth]{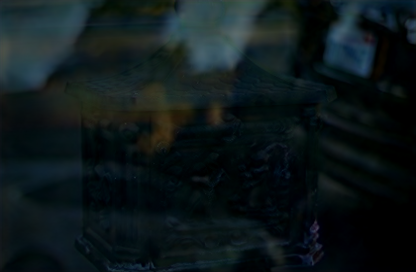}\vspace{2pt}
    \end{minipage}
    }
    \subfigure[DURRNet]{
    \begin{minipage}[b]{ 0.23\textwidth}
    \includegraphics[width=1.0\linewidth]{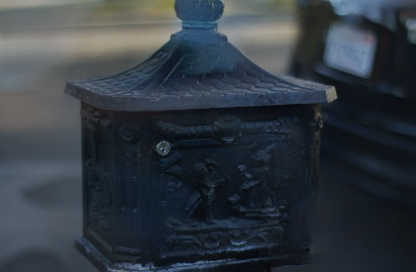}\vspace{0.1pt}
    \includegraphics[width=1.0\linewidth]{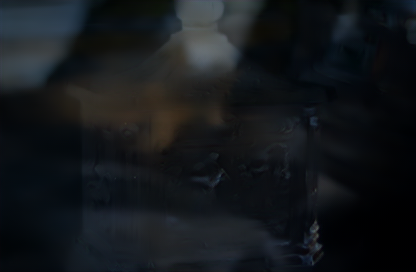}\vspace{2pt}
    \end{minipage}
    }
    \caption{Visualization of the different effects of  ProxNets and ProxInvNets in the proposed DURRNet.}
    \label{fig:ablation_Prox}
\end{figure}

From Table \ref{tab:ablation}, we can see that when ProxNets or ProxInvNets are removed from DURRNet there is approximately a 1 dB drop in PSNR. Therefore they are both essential components of the proposed DURRNet. To further visualize the functionality of ProxNet and ProxInvNet, Fig. \ref{fig:ablation_Prox} shows the single image reflection removal results of the DURRLayer w/o ProxNet, DURRLayer w/o ProxInvNet, and the complete model of DURRNet. We can see that when either ProxNets or ProxInvNets are disabled, the model can still produce relatively good results. This could be due to the network architecture design for ProxNet which includes a global skip connection, and ProxInvNet which adopts invertible networks as learnable transforms. In Fig. \ref{fig:ablation_Prox} (b) when ProxNets are disabled, the model would have difficulty to localize the reflection region, and in Fig. \ref{fig:ablation_Prox} (c) when ProxInvNets are disabled, the model would have difficulty at dealing with the contour regions of the reflections.

\noindent \textbf{The effectiveness of Multi-scale Architecture:} As shown in Fig. \ref{fig:DUSepNet}, the proposed DURRNet consists of $S$ scales of  DURRLayers to progressively estimate the transmission image and the reflection image from low-resolution scales to high-resolution scales. 
In Table \ref{tab:ablation} , we further analyzed the effectiveness of the multi-scale architecture. From the table, we can see that when the same total number of DURRLayer stages are fixed, \textit{i.e.}, $S \times K=8$, the proposed DURRNet (with $(S,K)=(4,2)$) achieves the best performance compared to other configurations, \textit{e.g.}, $(S,K)=(1,8)$ and $(S,K)=(2,4)$. This indicates that the multi-scale architecture can effectively and efficiently integrate information from different scales.

\section{Conclusions}
\label{sec:conclusion}

In this paper, we proposed a novel model-inspired single image reflection removal network named Deep Unfolded Reflection Removal Network (DURRNet). 
The proposed DURRNet is designed using deep unfolding technique and has clear interpretation. The image formation model and priors have been explicitly embedded into the design of the DURRNet architecture. Within each DURRLayer, ProxNets are used to model natural image priors and ProxInvNets which are constructed with invertible networks are used to impose the exclusion prior.
From experimental results, the proposed DURRNet is able to recover high-quality transmission and reflection images both quantitatively and visually.

\bibliographystyle{abbrv}
\bibliography{egbib}
\end{document}